# Detecting Special Nuclear Material Using Muon-Induced Neutron Emission

Elena Guardincerri[1], Jeffrey Bacon[1], Konstantin Borozdin[1], J. Matthew Durham[1], Joseph Fabritius II[1], Adam Hecht[3], Edward C. Milner[2], Haruo Miyadera[1], C. L. Morris[1], John Perry[1,3], and Daniel Poulson[1]

[1]*Los Alamos National Laboratory, Los Alamos, NM, 87545 USA.*

[2]*Southern Methodist University, Dallas, TX, 75205, USA.*

[3]*University of New Mexico, Albuquerque, NM, 87131, USA.*

**Abstract.** The penetrating ability of cosmic ray muons makes them an attractive probe for imaging dense materials. Here, we describe experimental results from a new technique that uses neutrons generated by cosmic-ray muons to identify the presence of special nuclear material (SNM). Neutrons emitted from SNM are used to tag muon-induced fission events in actinides and laminography is used to form images of the stopping material. This technique allows the imaging of SNM-bearing objects tagged using muon tracking detectors located above or to the side of the objects, and may have potential applications in warhead verification scenarios. During the experiment described here we did not attempt to distinguish the type or grade of the SNM.

# Introduction

Imaging with cosmic-ray muons has been the topic of a considerable recent body of work [1] because of the potential application to a wide range of difficult radiographic problems [2-11]. Cosmic-ray radiography makes use of muon scattering angles to measure the areal density along the muon's path. With muon tracking detectors on opposite sides of an object, an ensemble of trajectories can be used to generate tomographic images of the object's internal structure. Multiple scattering radiography enables threat detection in complex cargo scenes in ~minute time scales, using only the natural flux of cosmic-ray muons [12]. As this technique adds no artificial radiation dose, it is ideal for border protection and cargo scanning, where humans would potentially be subject to large exposures from more conventional x-ray radiography techniques.

In this report, we explore a new technique that neutrons generated via interactions with cosmic-rays to detect the presence of fissile material. It is well established that cosmic ray muons which stop in fissile material can induce a detectable amount of neutron emission [13-16]. We show that cosmic-ray-induced neutrons from a uranium target can be used to tag the cosmic-ray tracks which impinge on the target and produce images of the volume of fissile material. The result is an effective method for verifying the presence of fissile material, without revealing precise details on the exact amount or isotopic composition. The insensitivity of the technique to details make it potentially useful as a warhead verification technology, in a scenario where parties to a treaty agree to demonstrate the presence or absence of weapons components without revealing potentially sensitive aspects of weapons designs.

# Methodology

We describe the new method that we developed to tag cosmic-rays and use them as a probe of nuclear materials. Neutrons generated by secondary cosmic rays are used to tag cosmic-ray tracks. The tracking information from these events is used to image the source volume.

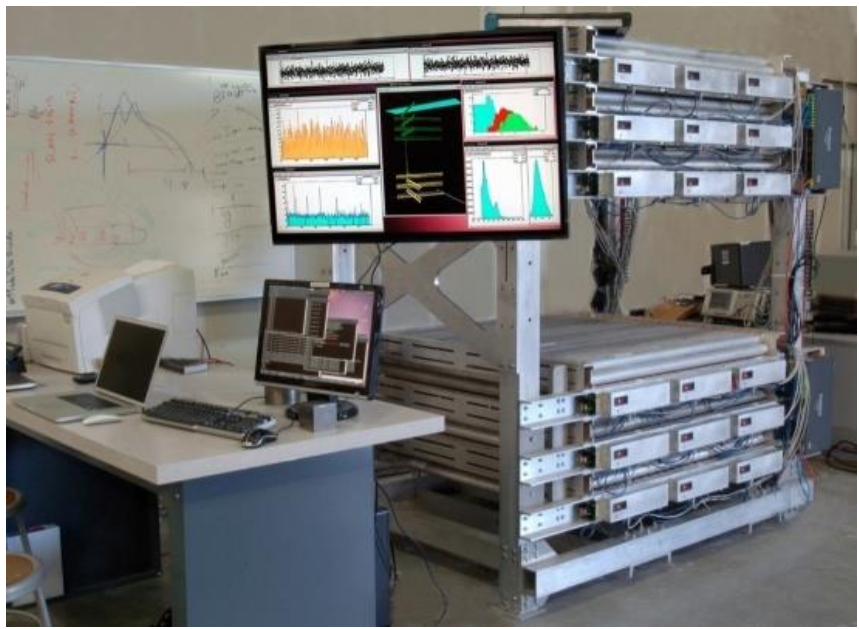

Figure 1 - Photograph of the mini muon tracker (MMT). The MMT consists of an upper and a lower tracker. Each tracker has 12 planes of drift tube detectors, the detector orientations were crossed between planes to provide tracking in both horizontal directions. Objects for study were placed in the approximate two-foot (60 cm) gap between the two detector "supermodules".

Radiography can take advantage of both the intensity and the direction of the cosmic-rays [17]. Tracking detectors above and below an object allow cosmic-rays which interact in the object material to be detected and used for image reconstruction. We have used the apparatus shown in Figure 1 [18] to measure stopped and transmitted cosmic-rays in an object placed between the two detector modules. The detector consists of a 576 drift tubes arranged in planes. Each of the two modules in Figure 1 consists of 6 planes of drift tubes, three of which are oriented along the horizontal X direction, and three of which are oriented along the horizontal Y direction, perpendicular to X. Each module can independently track cosmic rays and the trajectory information can be used to generate a focused transmission image at any distance from the detector.

A trajectory is defined by its coordinates of intersection with a plane and its direction cosines. Trajectories will put an object into focus if they are projected to the plane where the object is located.

Conceptually, the stopping length, $\lambda$, of cosmic-rays in material is inversely proportional to the stopping rate and can be related to the energy spectrum, $dN(E)/dE$, as

$$\frac{1}{\lambda} = \frac{dN}{Ndx} = \frac{1}{N}\frac{dN}{dE}\frac{dE}{dx} .$$

The majority of cosmic rays in the atmosphere are muons and electrons [19, 20]. The flux of muons at sea level is approximately 1 cm$^{-2}$min$^{-1}$, and the electron flux is about 35–40% of the muons. A small hadronic component, consisting of mainly protons and neutrons, is also present and it amounts to ~10% of the total flux at the 2200 m elevation of Los Alamos, where the measurement was conducted.

A plot of the energy spectrum for overhead muons at sea level is shown in Figure 2. The energy loss can be calculated using the Bethe-Bloch formula [21-23]. Over a wide range of momentum, the energy loss for cosmic-ray muons varies only logarithmically with momentum and is approximately proportional to the electron density, $Z/A$, where Z is the atomic number and A is the atomic mass. For dense material $\lambda$ is short compared to the muon decay length, $l = \beta c \gamma \tau$, where $\beta$ and $\gamma$ are the usual relativistic kinematic quantities, $c$ is the velocity of light, and $\tau = 2.2\ \mu s$ is the muon lifetime.

The stopping rate for muons is given by the density divided by the stopping length, and scattering is proportional to the square root of the density divided by the radiation length. Both the transmitted flux and the stopped flux provide radiographic signatures. For objects that are thin compared to $\lambda$, measuring the stopped flux typically provides smaller statistical uncertainty than the scattering signal.

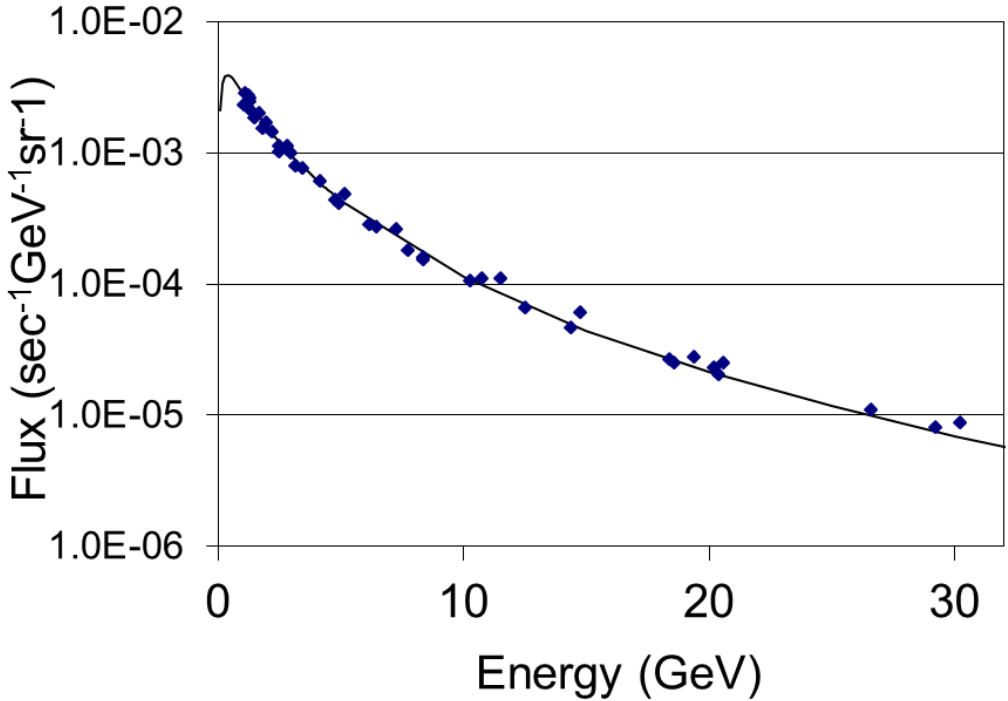

Figure 2 - Spectrum of vertical cosmic-ray flux at sea level. Solid symbols are the data [24]. The line is a parameterization.

Cosmic ray muons which encounter fissile material may induce neutron emission through three different mechanisms: photo-emission, muon-induced fission, and neutronic gain. As the highly relativistic muons pass through high-Z material, they will lose energy by emitting bremstrahlung photons. Prompt photoneutrons may be liberated when the heavy nuclei interact with these photons.

Muons which lose sufficient energy will come to rest inside the object. A positive muon ($\mu^+$) at rest will decay into a positron and two neutrinos, $\mu^+ \rightarrow e^+ + \nu_\mu + \overline{\nu_e}$. However, when a negative muon encounters the nuclear Coulomb field, it may be captured into an atomic orbital and rapidly de-excite into the ground state, emitting a set of high-energy muonic x-rays. In light nuclei, negative muons can also decay as a free particle, $\mu^- \rightarrow e^- + \nu_e + \overline{\nu_\mu}$, but when they are stopped in a material with high atomic number (Z), they can be captured by a bound proton and produce a neutron and a neutrino, $\mu^- \otimes A \to n + (A-1)^* + \overline{\nu_\mu}$. The lifetime of a muonic atom depends on its atomic number and was measured to be 71.6 ± 0.6 ns for $^{235}$U and 77.2 ± 0.4 ns for $^{238}$U[25].

In fissile material, neutrons liberated via photo- or muon-induced fission processes can trigger fission chains, leading to the emission of several neutrons, depending upon the effective multiplication of the neutronic system. Each fission reaction produces several gamma rays, 2-3 neutrons, and fission fragments. The time scale for neutron emission between subsequent fission events is typically on the order of ~10ns, so these secondary fission neutrons will be delayed relative to the primary neutron.

We note here that an additional source of neutrons may be present due to the protons that make up ~10% of the cosmic ray flux. These high-energy protons can also liberate neutrons from atomic nuclei through spallation reactions. The produced spallation neutrons are "prompt" with respect to the incoming protons and their multiplicity is proportional to the energy of the incident protons[26]. As our tracker cannot discriminate between proton and muon tracks, these will also be present in our data.

For any of these sources, the products of this muon-induced neutron emission create a distinctive signature that can be coupled to muon trajectories. Cosmic ray tracks which point to fissile material will be correlated with neutron emission, and can be used to create a tagged image within a coincidence window.

The amount of fast and thermal fission events depends upon the geometry of the object, the quantity of fissile material, as well as other materials that may reflect, absorb, or moderate neutrons, and it is described by the Boltzmann transport equation [27] through a series of losses and gains in the neutronic system.

There is a practical challenge in measuring signals from fast or thermal fission. A fast signal resulting from muon-induced fission consists of neutrons and gammas emitted in a narrow time window on the order of hundreds of

LA-UR-14-21076

nanoseconds. The challenge with this energy region is that the mean free path of a neutron in uranium is large (order of centimeters). On the other side, if neutrons from thermal fission are to be measured, the timing coincidence window must be extended by hundreds of microseconds to account for the moderation time. The wider coincidence gate results in higher rate of accidental coincidences due to random background.

**Experiment**

Data were taken with the Los Alamos Mini-Muon Tracker (MMT) [18] imaging a 19 kg (1000 $cm^3$) low enriched uranium (LEU) cube (19.7% $^{235}$U) . The LEU cube was mounted on a wooden platform on top of the MMT lower "supermodule". The cube was imaged using muon scattering, muon transmission, and muon absorption techniques, as well as using neutron tagging in coincidence with muon scattering or absorption.

Muon-induced neutron emission events in the LEU cube produced neutrons that were detected in two 12.5 cm diameter by 5 cm deep EJ301 liquid scintillator detectors. The neutron detectors were mounted side by side 17.5 cm from the center of the LEU cube as shown in Figure . The liquid scintillator signal has two light decay time constants that can be used to discriminate signals due to gammas (fast) and neutrons (slow). The Mesytec NIM module MPD-4 was used to discriminate between gammas and neutron signals using Pulse Shape Discrimination (PSD) techniques, as described in [28]. In particular, the module was calibrated using fission neutrons from a $^{252}$Cf source and gammas from a $^{137}$Cs source and its neutron-trigger output was fed to a dedicated input on the MMT digitizing boards. The times of neutron signals from the detectors were recorded in the data stream along with the drift tube data.

Measurements were performed for 48 hours, with a measured fluence of $3.33 \cdot 10^5$ cosmic-rays interacting with the LEU. From this measured fluence, $1.04 \cdot 10^5$ particles stopped in the LEU, and the rest of the flux was scattered to the lower detector. The majority of stopping particles are likely electrons: the electron flux is in fact about 35–40% of the muon flux. At the 2200 m elevation of Los Alamos the integrated muon flux is a factor of approximately 1.5 than that of sea level. The larger muon flux at altitude is consistent with our measurements.

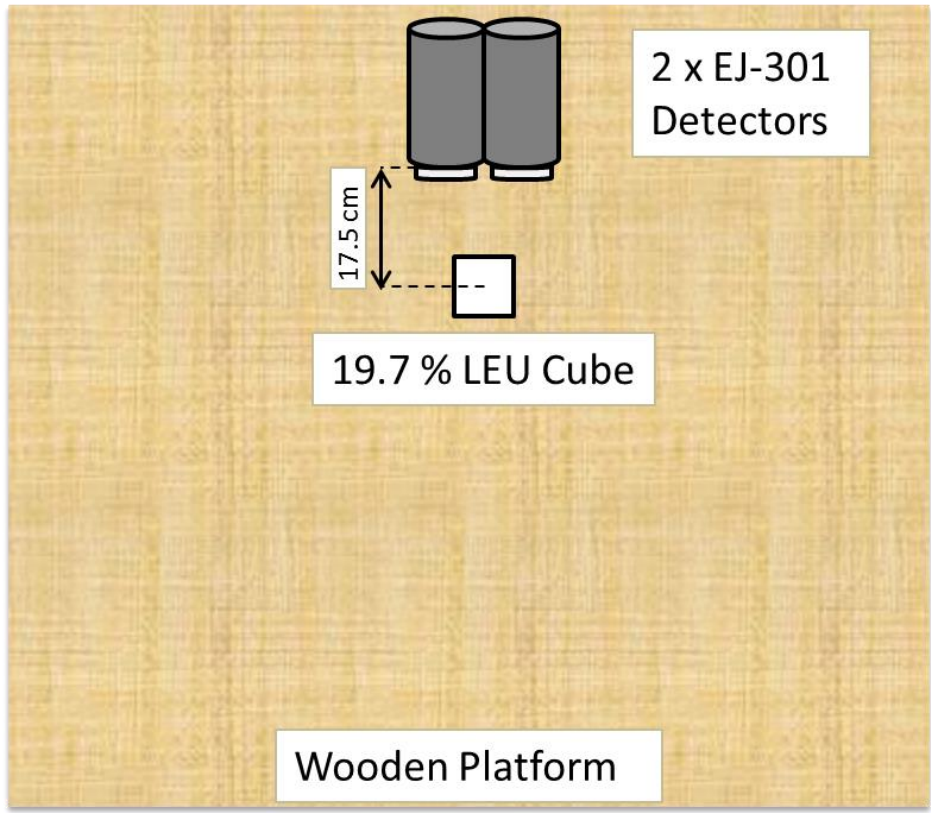

Figure 3 – Experimental configuration of EJ-301 liquid scintillator detectors and the LEU cube target. The active region of the detectors is located 17.5 cm from the center of the LEU cube target. Both the detectors and uranium cube sit on a wooden platform located between the MMT supermodules

LA-UR-14-21076

# Results

In Figure 3 we show the four different types of imaging performed with the MMT: multiple scattering, transmission, stopped tracks, and neutron tagged tracks in coincidence with stopped tracks.

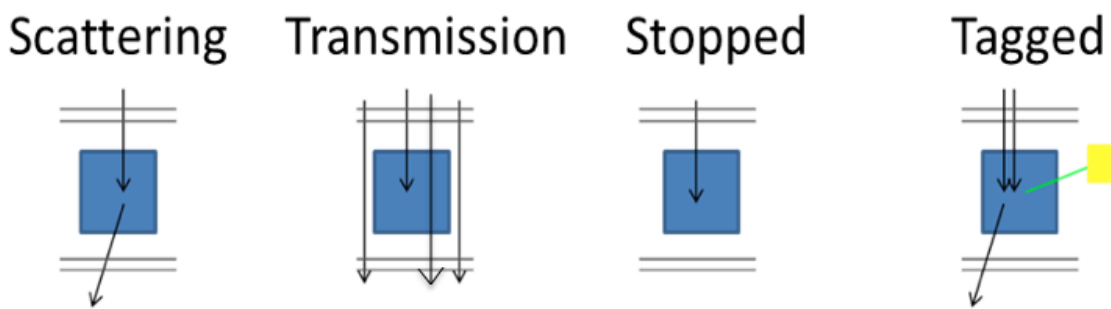

Figure 3 –Illustration of four types of radiography possible with muons. The tagged data are a sum of tagged stopped and transmitted events

The timing distribution between the incoming muon track and neutron emission is shown in Figure 4, for tracks which pass through the cube and tracks that are stopped in the uranium. The signal-to-noise ratio is more than 100:1 averaged over the entire field of view of the tracker (~1.2 m×1.2 m). Approximately half of the neutrons which are in coincidence with stopped tracks are emitted with a time constant of $85 \pm 3.3 \, ns$, while the neutrons correlated with through tracks show no such tail.

Cosmic rays which pass completely through the cube generate prompt primary photoneutrons and secondary neutrons from fission chains in the LEU. For the stopped tracks, both of these mechanisms are present, along with muon-induced fission neutrons from captured negative muons. These neutrons can also induce secondary fission, leading to a time between muon capture and secondary neutron emission on the order of the lifetime of the muonic U atom plus the time scale for neutron multiplication. We therefore attribute tail of the stopped muon timing distribution to muon-induced fission and secondary fission in the low enriched uranium cube.

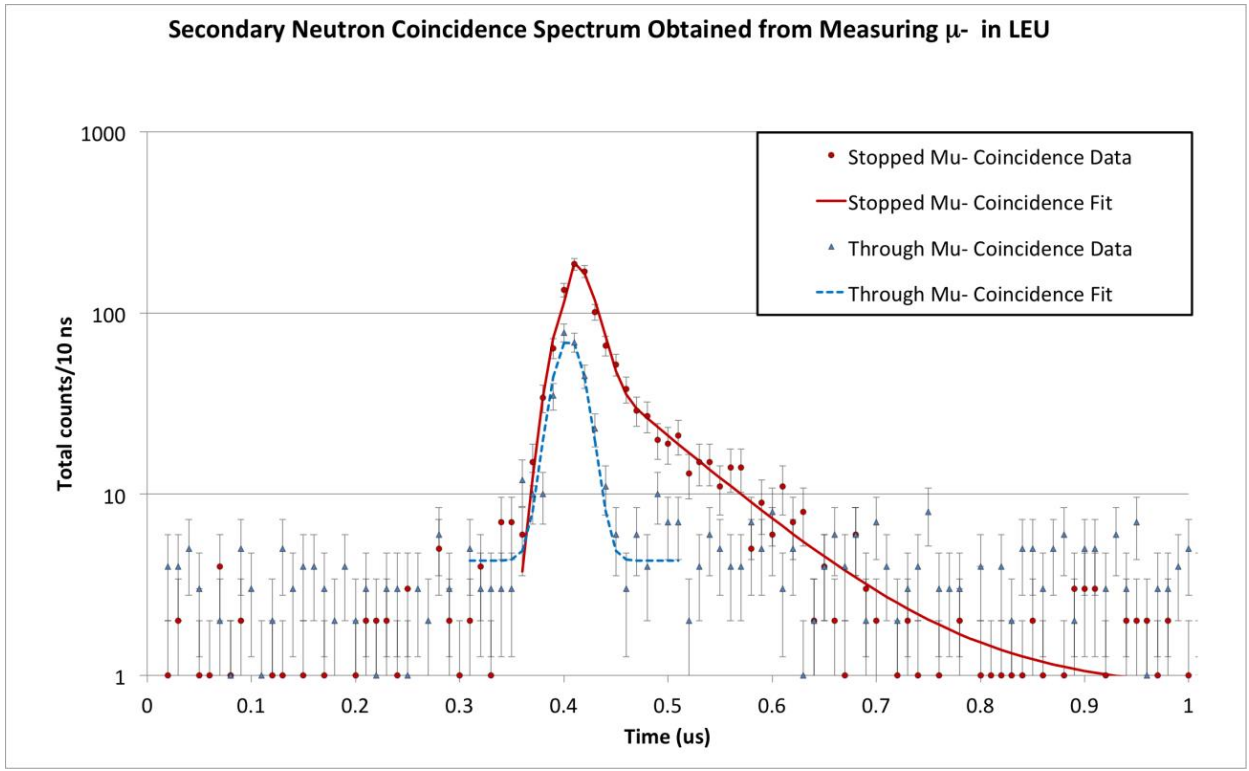

Figure 4 - Spectrum of coincidence time between neutrons and muons. On the vertical axis the number of recorded coincidences per 10 ns horizontal bin is shown. The time signal for the muons is provided by the time zero from the track fitter. The plot shows the difference in the detected neutron time and the muon time for stopped trajectories and through trajectories. A log likelihood fit was used to fit the data. The exponential time constant of this fit was $85 \pm 3.3\ ns$, which is near the measured lifetime states of muonic uranium found in literature.

The image reconstructed from neutron-tagged events can be assessed on different planes between the detector supermodules. The source of the neutrons is assumed to be within the plane with the sharpest reconstructed image. This plane is chosen to be the same between all four methods of imaging. A set of images from all four methods of muon radiography is shown in Figure 5. The neutron tagged image (Figure 5, right column) is seen to have a signal-to-noise ratio that matches the quality of the scattering image, although the statistics and position resolution are not as good.

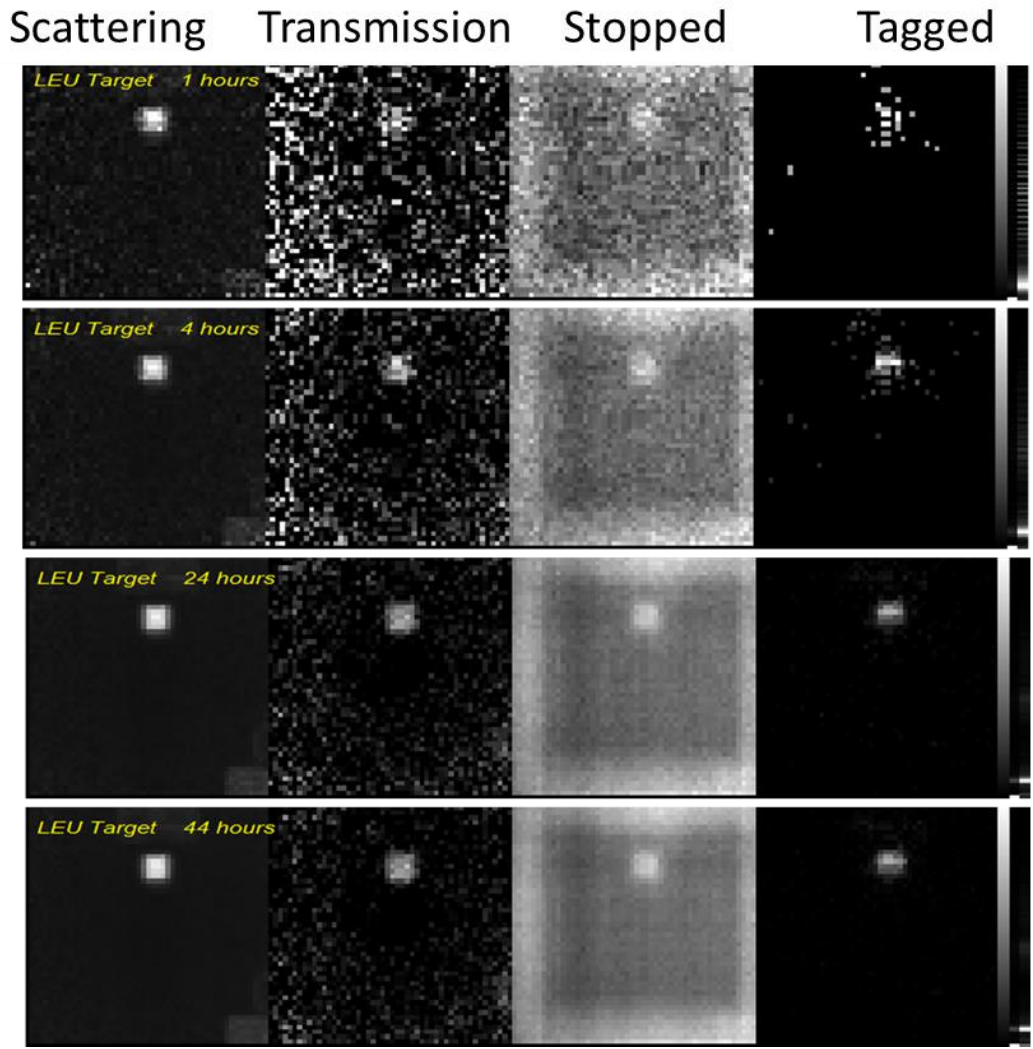

Figure 5 - Slices from muon radiographs showing four kinds of radiography possible with cosmic-rays. The object is a 19 kg cube of low enriched uranium. The scale for each of the grey scale images is linear where black is zero. Exposure times were varied (vertical direction) from 1 to 44 hours.

Scattering provides good images within minutes. By one hour the size and shape of the object is clear. With neutron tagged tracks the object is detected with high reliability in an hour, but an accurate determination of the shape requires tens of hours. For fissile materials, the neutron signal is largely enhanced.

This technique for imaging fissile material may potentially be useful for warhead verification. In such a scenario, two parties may wish to demonstrate the presence or absence of fissile material in a warhead without exposing details

LA-UR-14-21076

on the exact composition or configuration. As this technique requires long exposures to extract precise images, the count time may be limited to keep spatial resolution to a mutually acceptable level. Additionally, the exact isotopic composition of the target is not accessible with this technique.

## Conclusion

In conclusion, we have shown a proof-of -principle measurement that demonstrates a new technique for imaging fissile material with a combination of cosmic ray tracking detectors and neutron detectors. Cosmic ray tracks which are stopped in LEU induce neutron emission on a time scale which is consistent with muon-induced fission and subsequent neutron gain. These tracks can subsequently be used to image the LEU target. Future work will more fully explore the usefulness of this technique for possible applications in treaty verification efforts.

## Acknowledgements

This work was supported in part by the United States Department of State, and the Defense Threat Reduction Agency of the United States Department of Defense, but it does not necessarily reflect the views or position of the U.S. government on the issues discussed herein.

LA-UR-14-21076